\documentclass[preprint,superscriptaddress]{revtex4}

\usepackage[dvips]{graphicx}
\usepackage{amssymb,amsfonts,amsmath}
\begin{document}
\begin{center}
{\Large \bf Supporting Information}\\
{\large \bf ``Dynamics of the bacterial flagellar motor with multiple stators''}\\
\vspace{0.5cm}
Giovanni Meacci and Yuhai Tu\\
%$^{*}$\\
%\vspace{0.3cm} IBM T. J. Watson Research Center\\ P.O. Box 218, Yorktown
%Heights, NY 10598\\
%$^*$Corresponding author (email: yuhai@us.ibm.com)\\
\end{center}
%\vspace{2.5cm}

\section{Torque-speed curve measurement}
The measurement of the torque-speed curve is usually done by fixing the cell to a glass slide and tethering a polystyrene bead to the flagellar hook. An optical trap monitor the rotational speed of the bead and the motor torque is calculated from $\tau=(\xi_{L}+\xi_{R})\omega\approx \xi_{L}\omega$, where $\xi_{R}$ is the drag coefficient due to the internal friction in the motor, $\xi_{L}$ is the bead drag coefficient, and $\omega$ is the angular velocity. The torque-speed curve is then obtained changing $\xi_{L}$ by varying the bed size \cite{SHHI03} or the viscosity of the external medium. An alternative method is to tether a cell to a glass coverslip by one of its shortened flagellar filament and expose the cell to a rotating electric field \cite{BT93}. Then the motor is broken spinning the cell backward. The difference between the cell body speeds before and after the broke of the motor, at the same value for the external applied torque, is proportional to the motor torque.

\section{Hook spring compliance}
Fig.~S1 shows the compliance curve $F$ (motor torque versus angular displacement $\theta-\theta_{L}$) similar to the experimental one \cite{BBB89}. The following expression is used in our simulations:
\begin{eqnarray}
F(x)=\kappa_{0} x+\Delta x(\kappa_{1}-\kappa_{0})
%{}{}
%\nonumber\\
log[1+exp\big((x-x_{0})/\Delta x\big)], \label{qg2}
\end{eqnarray}
with $x\equiv \theta-\theta_{L}$, $\kappa_{0}$ the spring constant a low load, $\kappa_{1}(\gg \kappa_0)$ the spring constant a high load. At both low and high loads the spring behaves linearly with different spring constants $\kappa_0$ and $\kappa_1$ respectively. $\Delta x$ is the angular displacement interval of the non linear region centered around the turning point $x_{0}$. For the values of these parameters see Tab.~S1.

\section{Distribution functions of $t_{m}$ and $t_{w}$}
Fig.~S2 shows the distribution functions for $t_{m}$ and $t_{w}$ at different values of the load for the case N$=$1. Fig.~S2(a) shows the average waiting and moving times. The arrows labeled with the letter (b), (c), and (d) indicate the points (i.e. the speed values) where the probability distributions for $t_{m}$ and $t_{w}$ are shown in Fig. S2 (b), (c), and (d) respectively. The averaged waiting-time decreases slightly with the speed, while the averaged moving-time decreases by four orders of magnitude. The two time scales crossover at a speed around 170Hz, which naturally defines two regimes: i) $\langle t_{w}\rangle \ll \langle t_{m}\rangle$ and ii) $\langle t_{w}\rangle \gg \langle t_{m}\rangle$. The crossover speed corresponds roughly to $\omega_{n}$.

Figs.~S2(b), (c), and (d) show the differences in the distribution functions for $t_m$ and $t_w$ at high, medium, and low load respectively. The waiting times are exponentially distributed because the waiting time interval is determined by independent chemical transitions, i.e., by Poisson processes. The average waiting time thus depends on the stator jump rate $k$, which varies between two constants $k_{+}$ and $k_{-}$ (except for extreme high load where $k=0$): $\langle t_w \rangle\propto 1/k$. This explains why the averaged waiting time only weakly depends on the load.  At low load, premature jumps are rare and the average angular movement $\delta_m$ is determined by a single stator jump and has a peaked distribution centered around $\delta_0/N$. This explains the peaked distribution for $t_m$ at low load, as shown in Fig.~S2(b). At medium load, both $\langle t_{m}\rangle$ and the value of $ t_{m} $ corresponding to the peak of the distribution increase, as shown in Fig.~S2(c). At high load, the $t_m$ distribution develops a flat region for shorter time intervals, as shown in Fig.~S2(b). This is a consequence of the decreasing slope of the total potential felt by the rotor on the positive force side. As a result, fluctuations of the rotor angle $\theta$ due to thermal noise increase, which leads to many short moving time intervals.

\section{Dependence of the torque plateau region on k$_{+}$ and $\delta_{c}$}
In order to understand the origin of the torque plateau region, we have studied the dependence of $\omega_{n}$ on the ratio $r \equiv k_{+}/k_{-}$ and on the cutoff $\delta_{c}$ for the case N$=$8 (similar results have been obtained for different values of N). We define $\omega_{n}$ as the speed value at which the torque decreases 10$\%$ from its value at stall. The size of the plateau regime is characterized by the following quantity:
\begin{eqnarray}\label{shoulderApp}
\Sigma= \omega_{n}/\omega_{max}.
\end{eqnarray}
Fig.~S3(a) shows the torque-speed curves for N$=8$ and for two different values of $r$: $r=0$ and $r=1.2$. For $r=0$ the torque, after a small plateau, decreases linearly with the speed. Increasing $r$, $\omega_{n}$ increases and the torque-speed curve increase its concavity. In Fig.~S3(b), $\Sigma$ increases from 0.3 to 0.6 as $r$ increases from 0.05 to 1. This corresponds to the values of $r$ that can produce a well defined shoulder and at the same time maintain the independence of $\omega_{max}$ on N, i.e. $|\Delta|<1$ (see Fig.~4(b) in the main text).

Fig.~S3(c) shows four torque-speed curves for different values of the cutoff $\delta_{c}$. Starting from the value used in the main text, i.e. $\delta_{c}=\delta_{0}$, $\omega_{n}$ increases with $\delta_{c}$. In particular, $\Sigma$ increases significantly from $\delta_{c}=\delta_{0}$ to $\delta_{c}=3 \delta_{0}$, at which the plateau size $(\omega_n)$ reaches a value slightly bigger than $60 \%$ of the maximum speed (Fig.~S3(d)).

In conclusion the extension of the torque plateau region increases with the value of rate $k_{+}$ and the cutoff $\delta_{c}$, in consistent with our theory.

\section{Robustness of the results against different rotor-stator potentials and load-rotor forces}
In order to verify the independence of our results on the specifics of the rotor-stator potential, we studied our model with asymmetric potentials and a smoothed symmetric potential with a parabolic bottom (instead of the V-shaped bottom). In the asymmetric case, the slope $\tau_+$ of the left branch of the potential is much smaller than the slope $\tau_-$ of the right branch, similar to the potential used in \cite{XBBO06} (see the values used for $\tau_{+}$ and $\tau_{-}$ in Tab.~S1). Our model with the asymmetric potential yields qualitatively similar results as with the symmetric potential used in the main text. In particular, the maximum speeds near zero load are independent of the number of stators (see Fig.~S4) provided the stator jumping rates satisfy: $k_{-}/k_{+}\gg \tau_{-}/\tau_{+}$. Such a requirement can be understood intuitively in the following way. Given the condition $\tau_{-} \gg \tau_{+}$, the force equilibrium (waiting phase) is achieved by having one stator spending part of its time dragging the rotor while all the other stators are pulling the rotor. The waiting period ends when this dragging stator jumps with a rate that depends on the fractions of time it spends on the two sides of the potential, which depend on the ratio $\tau_-/\tau_+$. Therefore, the condition that the maximum speed is dominated by $k_-$ (instead of $k_+$) has to be weighted by the ratio $\tau_-/\tau_+$.

We have also studied a ``semi-parabolic'' potential $V^{P}(\Delta\theta\equiv \theta-\theta_{L})$ (see insert in Fig.~S5) defined as:
\begin{equation}
V^{P} = \left\{
\begin{array}{ll}
\tau_{0} [|\Delta\theta| - (\Delta\theta_{0}/4)] & \textrm{if $|\Delta\theta| > \Delta\theta_{0}/2$},\\
\tau_{0} \Delta\theta^{2}/ \Delta\theta_{0}  & \textrm{if $ |\Delta\theta| < \Delta\theta_{0}/2$},
\end{array}\right.
\end{equation}\label{parabolic}
where $\tau_{0}$ is the positive slope of the symmetric potential, and $\Delta\theta_{0}$ is the angular interval of the parabolic region centered around the bottom of the potential. Correspondingly, the chemical rate is a continuum function of $\Delta\theta$:
\begin{eqnarray}
k(\Delta\theta)=k_{+} +(k_{-}-k_{+})/
[1+exp\big(-(\Delta\theta)/\Delta\theta_{0}\big)] \label{kcontinuum}
\end{eqnarray}
Fig.~S5 shows the torque-speed curves for $N=1,2,...,8$. The curves show the same characteristics as for the V-shaped potential shown in the main text. The plateau region is a little wider. This is due to the change of the slope near the potential bottom. A lower value of the slope slows down the motion of rotor, increasing the premature jump probability before it reaches the bottom.

Next, we considered different forms of the force function $F$ between the load and the rotor.
Fig.~S6 shows torque-speed curves for two cases: with and without a spring between the load and the rotor for different ratio $k_+/k_-$. The case with spring between the load and the rotor is studied in the main text; the case without spring corresponds to infinite spring constant (rigid connection between load and rotor) with the following equation for the rotor:
\begin{eqnarray}\label{eqrotorApp}
\frac{d\theta}{dt}=-\frac{1}{\xi_{R}+\xi_{L}}\frac{\partial V}{\partial \theta}
+\sqrt{2k_{B}T / (\xi_{R}+\xi_{L})} \alpha(t).
\end{eqnarray}
Contrary to the model proposed in \cite{XBBO06}, the concavity of the torque-speed curve does not depends on the strength of the hook spring: the torque-curves are almost identical with and without spring. Instead, the concavity of the torque-speed curve depends on the ratio of the jump rates $k_+/k_-$, and also on the cutoff $\delta_{c}$ as shown before in Fig.~S3. In particular, as shown in Fig.~S6, for $k_{+}=0$ the concavity is zero, and it increases as the ratio $k_{+}/k_-$ increases.

\newpage
%\noindent
%{\bf Figure \& Table Legends}
%\vspace{0.5cm}
%\noindent

\begin{figure*}[!t]
\vspace{2cm}
\centerline{\includegraphics[angle=0,width=0.6\textwidth]{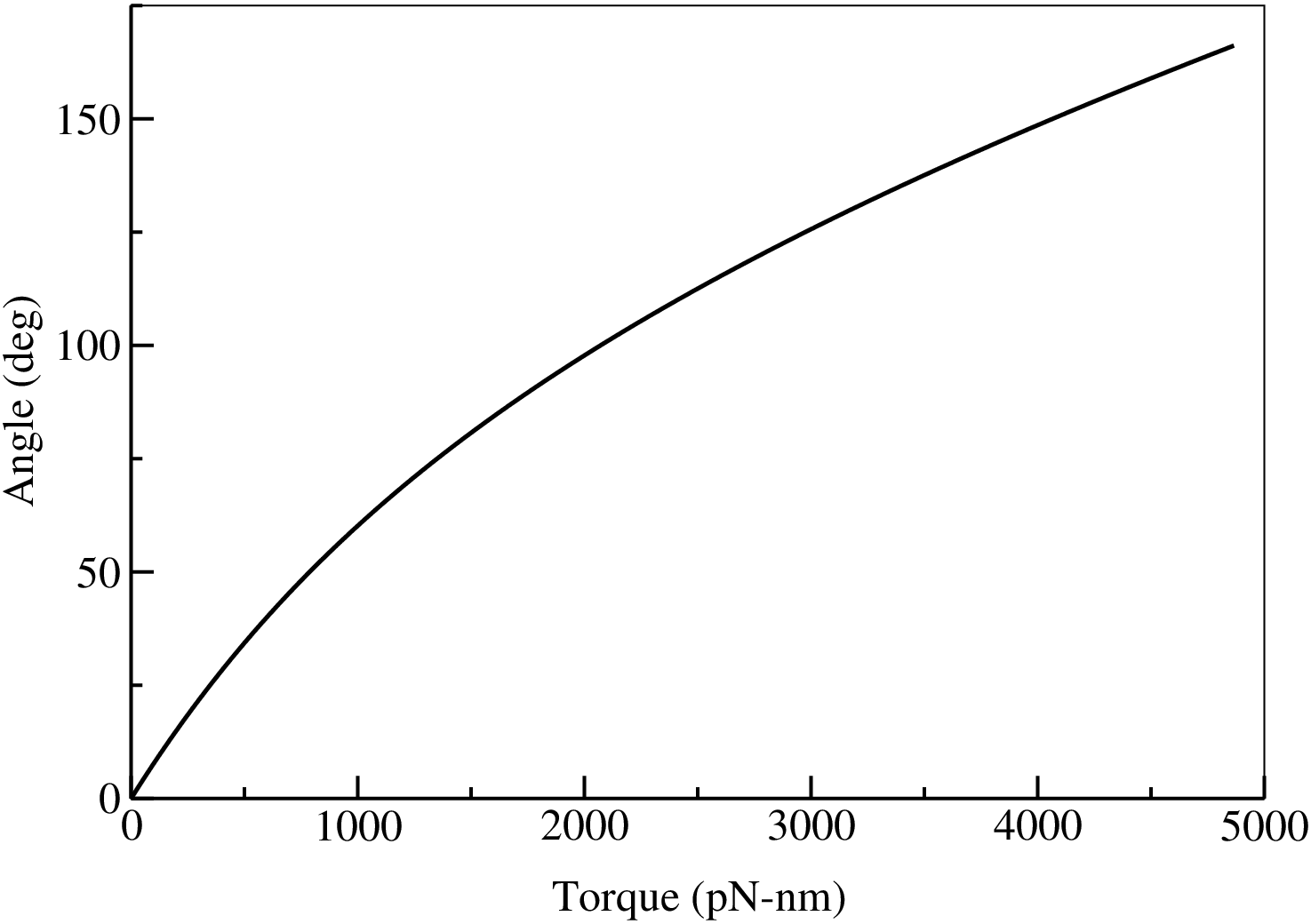}}
%{\bf Fig.~S1.} 
\caption{Compliance curve. The graph shows the angular displacement, $\theta-\theta_{L}$ , between the rotor and the load as a function of the rotor torque. The non linear spring behavior follows approximately the experimental measurement in reference \cite{BBB89}. The curve corresponds to the function $F(x)-F(0)$ used in our simulations.}
%\vspace{0.5cm}
\end{figure*}
\vspace{2cm}
\begin{figure*}
\vspace{2cm}
\centerline{\includegraphics[angle=0,width=0.6\textwidth]{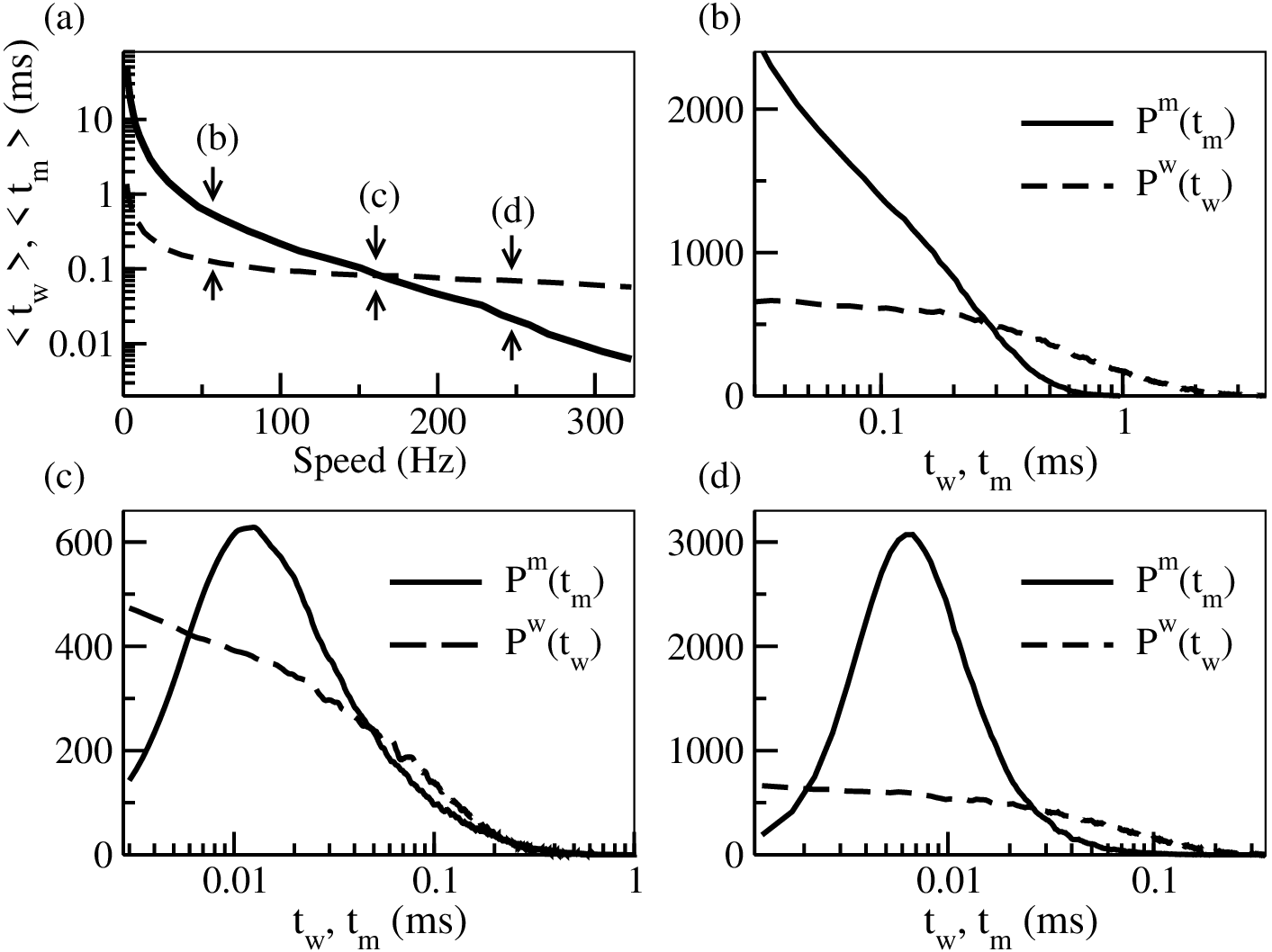}}
%\noindent
%{\bf Fig.~S2.} 
\caption{Waiting-time and moving-time statistic. (a) The average waiting-time (dashed line) and moving-time (solid line) averaged over 500 revolutions as a function of the rotational speed for N$=1$. With increasing speed, the averaged waiting-time slowly decreases from 1ms to 0.1ms while the average moving-time decreases much faster over four orders of magnitude, from roughly 50ms to 0.005ms. The vertical arrows labeled with the letters (b), (c), and (d), indicate the points where the $t_w$ and $t_m$ distributions are shown in the corresponding figures (b), (c), and (d). The value of the load is 14, 1, and 0.1pN-nm-s-rad$^{-1}$, for (b), (c), and (d) respectively. Probability distributions $P^{m}$ for $t_m$ are shown by solid lines and $P^{w}$ for $t_w$ are shown by dashed lines. The waiting-times are exponentially distributed in all load range. The moving-times show peaked distributions. The peak at high load is partially hidden by a flat region for small $t_w$, as a consequence of the rotor fluctuations.}
%\vspace{0.5cm}
\end{figure*}
\vspace{2cm}
\begin{figure*}
\vspace{2cm}
\centerline{\includegraphics[width=0.6\textwidth]{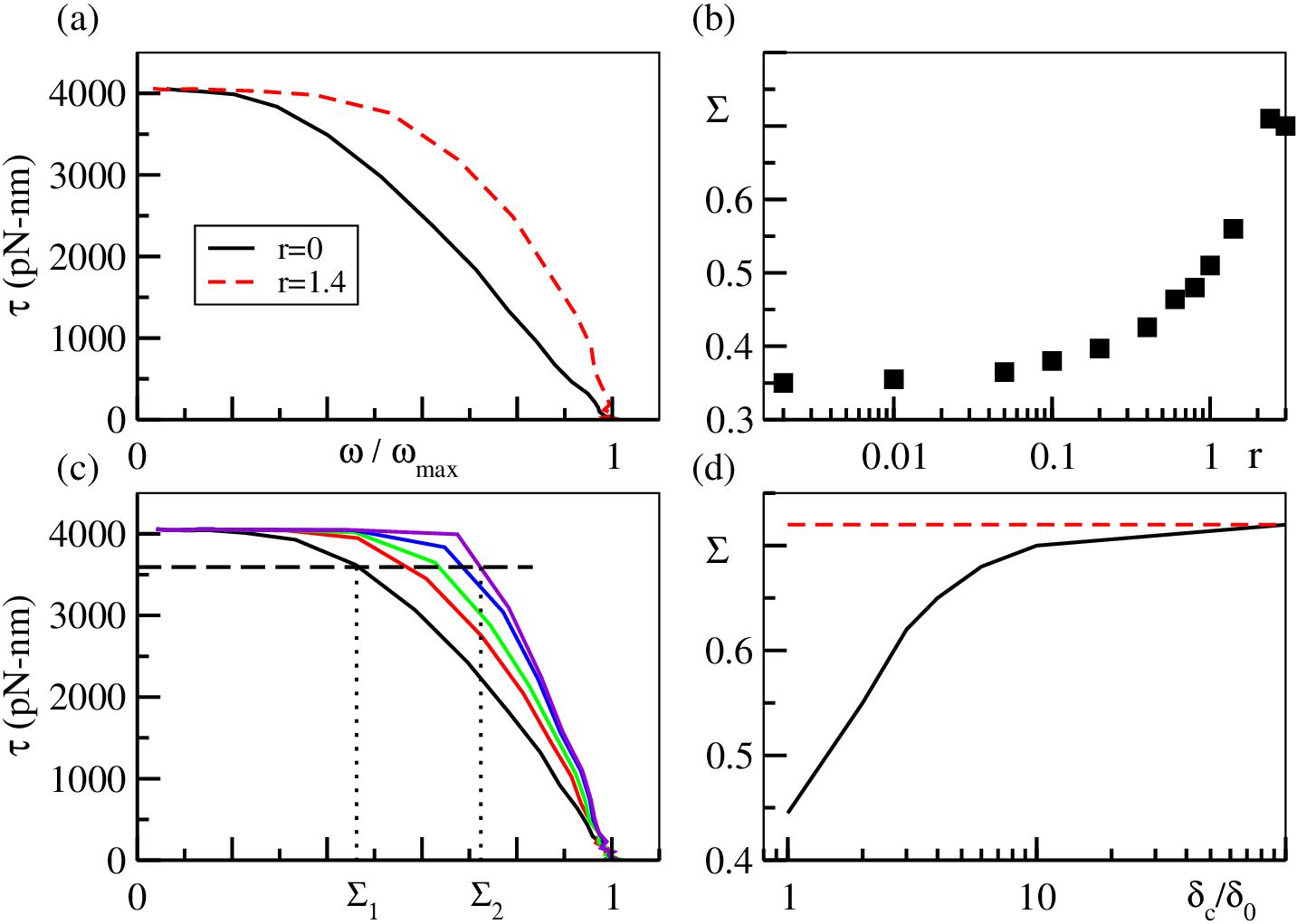}}
%\noindent
%{\bf Fig.~S3.} 
\caption{Dependence of the knee speed $\omega_{n}$ on $r\equiv k_{+}/k_{-}$ and on the cutoff $\delta_{c}$ for the symmetric potential with a parabolic bottom. (a) Motor torque as a function of the normalized angular speed for two different values of $r$:  $r=0$ (black solid line), and  $r=1.4$ (red dashed line). Different from the V-shaped potential, there is a small plateau regime even for $r=0$. (b) $\Sigma \equiv \omega_{n}/\omega_{max}$ as a function of $r$. (c) Torque-speed curves for different values of the cutoff $\delta_{c}$. Black, red, green, blue, and violet lines correspond to values of $\delta_{c}$ of 1, 2, 3, 6, 100$\delta_{0}$ respectively. $\Sigma_{1}$ and $\Sigma_{2}$ correspond to the values of $\Sigma$ for $\delta_{c}=\delta_{0}$ and $\delta_{c}=100 \delta_{0}$ respectively. (d) $\Sigma$ as a function of the ratio $\delta_{c}/\delta_{0}$ shown as the black solid line. The red dashed line corresponds to the asymptotic value $\Sigma (\delta_{c} \rightarrow \infty)$.}
%\vspace{0.5cm}
\end{figure*}
\vspace{2cm}
\begin{figure*}
\vspace{2cm}
\centerline{\includegraphics[width=0.6\textwidth]{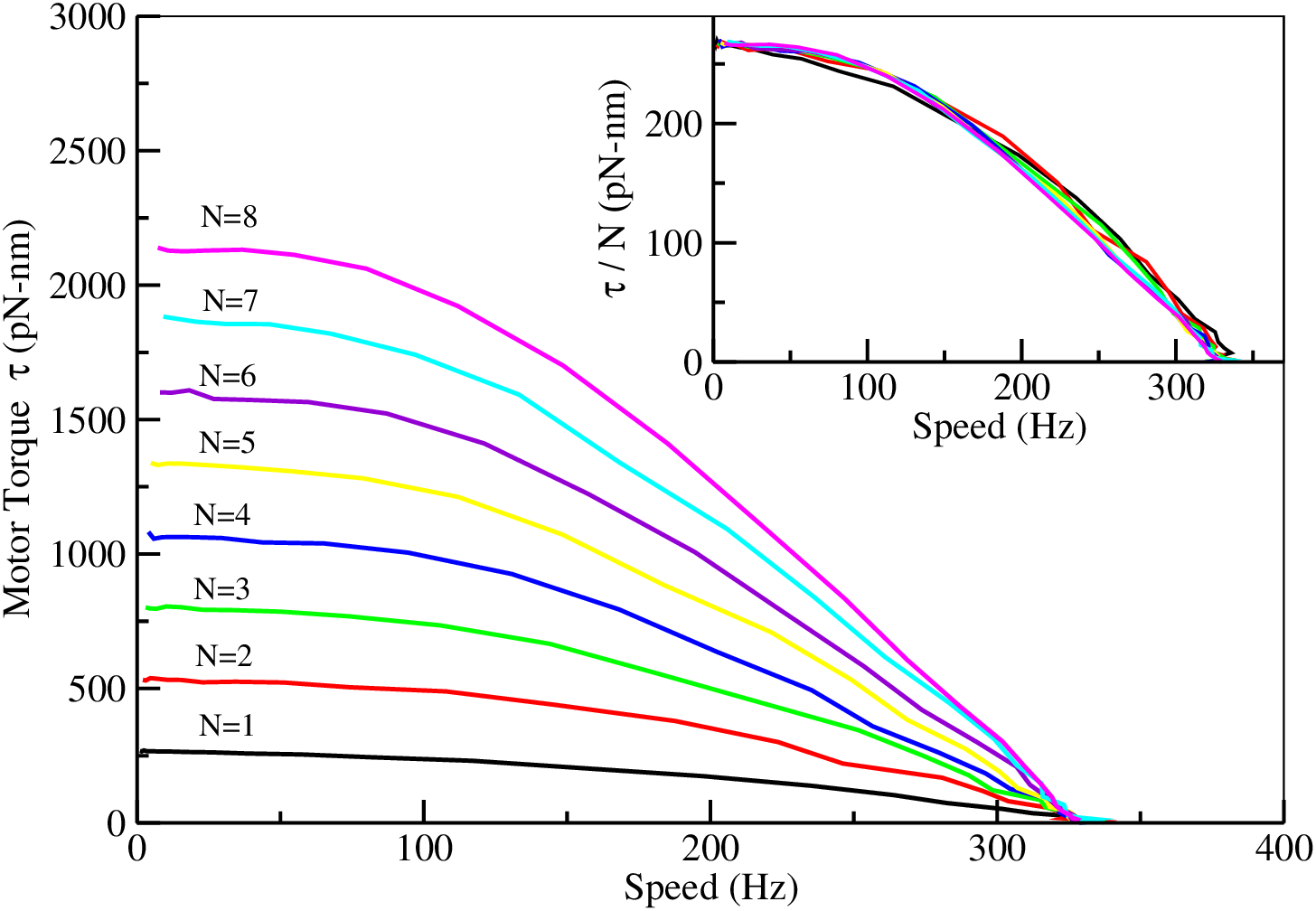}}
%\noindent
%{\bf Fig.~S4.} 
\caption{Torque-speed curves for an asymmetric potential. The torque-speed relationship is very similar to the symmetric potential case. The figure shows the motor torque $\tau$ as a function of rotational speed for N$=1, 2,...,8$. For a given value of N the torque is almost constant up to the knee before it decreases roughly linearly. At low speed, the motor torque increases linearly with N. At high speed near zero load all curves collapse to the same maximum speed. These characteristics are apparent in the insert, where the torque per stator $\tau/N$ versus $\omega$ are shown.}
%\vspace{0.5cm}
\end{figure*}
\vspace{2cm}
\begin{figure*}
\vspace{2cm}
\centerline{\includegraphics[width=0.6\textwidth]{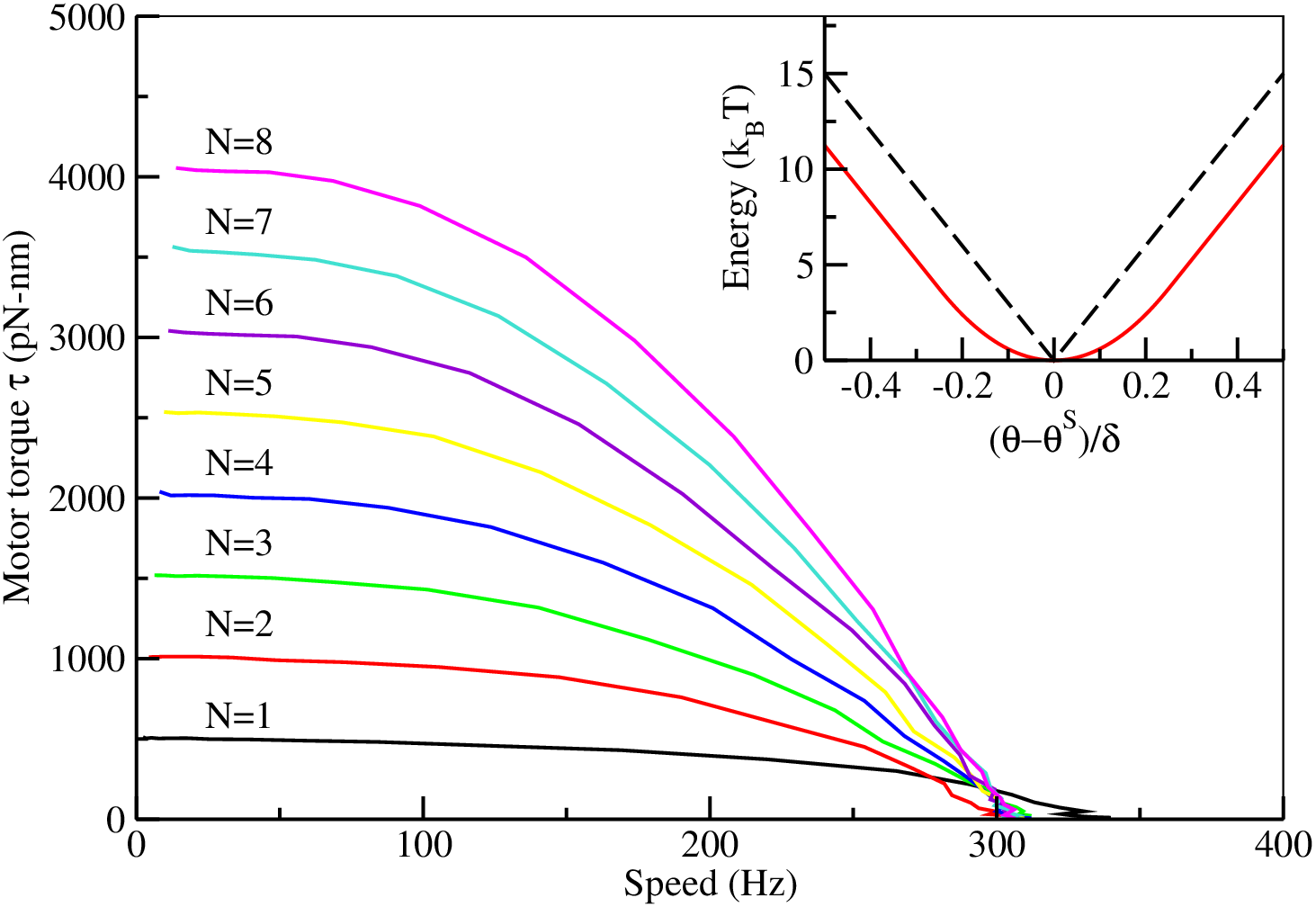}}
%\noindent
%{\bf Fig.~S5.} 
\caption{Torque-speed curves for the symmetric potential with a smooth, parabolic bottom. The behavior is very similar to those observed in the symmetric and asymmetric V-shaped potential cases. The figure shows the motor torque $\tau$ as a function of rotational speed for N$=1,2,...,8$. For a given value of N the torque is practically constant up to the knee, which is reached at almost the same speed value for different number of stators N, then it decreases linearly with $\omega$. The insert shows the ``semi-parabolic'' potential (solid red curve) and the symmetric V-shaped potential (dashed black curve) for comparison.}
%\vspace{0.5cm}
\end{figure*}
\vspace{2cm}
\begin{figure*}
\vspace{2cm}
\centerline{\includegraphics[width=0.6\textwidth]{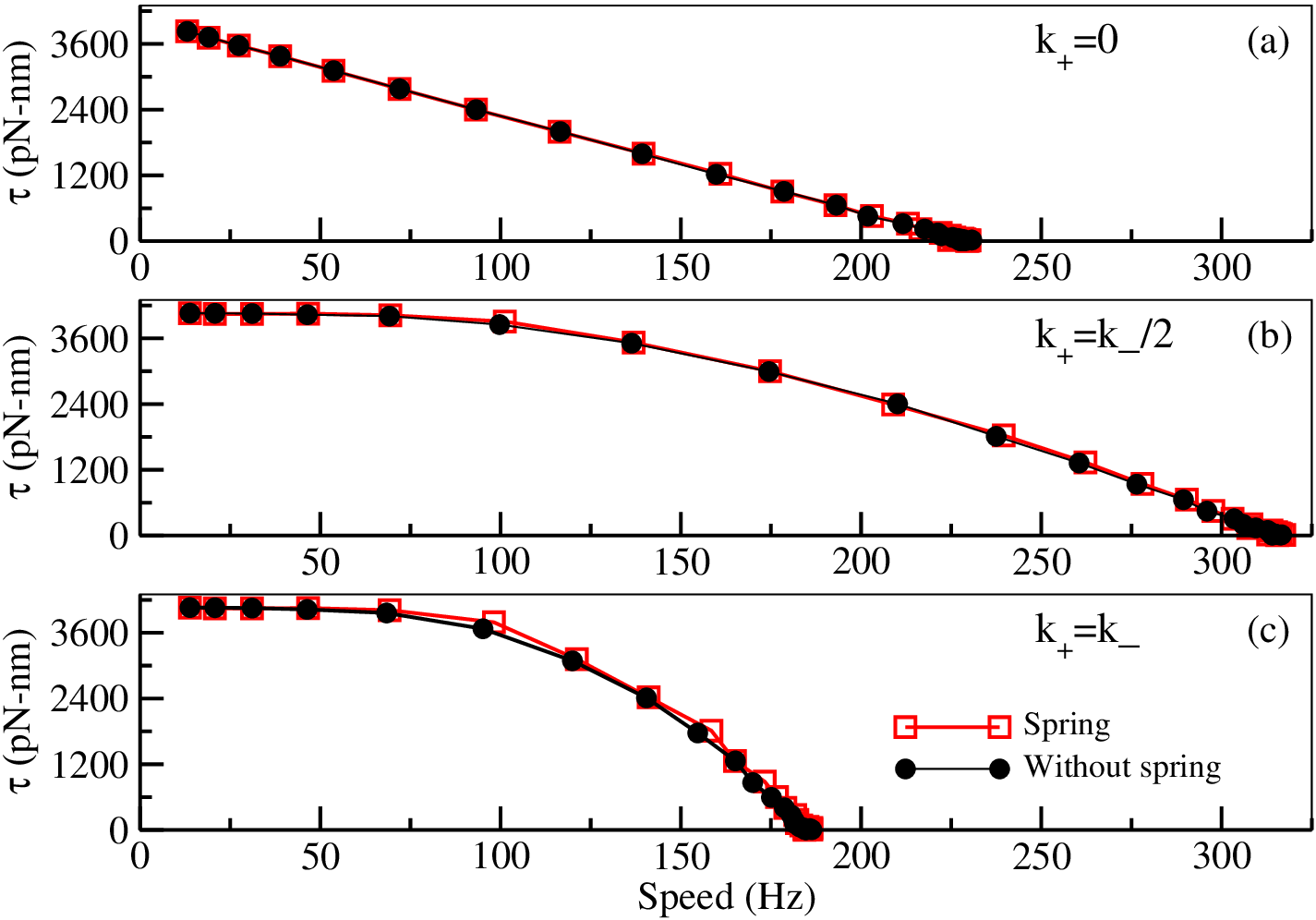}}
%\noindent
%{\bf Fig.~S6.} 
\caption{Independence of the torque-speed characteristics on the load-rotor spring. Torques-speed curves $(N=8)$ with spring (dotted lines) and without spring (squared lines) show almost identical behavior. Instead, the concavity of the torque-speed curve depends on the stator jump rate ratio $r=k_{+}/k_-$: (a) $r=0$, (b) $r=0.5$, and (c) $r=1$.}
%\vspace{5cm}
\end{figure*}
%\noindent
%{\bf Table~S1.} Parameters used in the calculation.
\begin{table}
\centering
%\begin{center}
\begin{tabular}{|c|c|c|}
  %$\lambda$&
  \hline
   Quantity & Value & Comment \\
\hline
  %$d_{_\lambda}$&
   $\xi_{R}$& 0.002pN-nm-s-rad$^{-1}$& Estimated from [2]\\
   \hline
   $\xi_{L}$&$\approx$ (0.002-50)pN-nm-s-rad$^{-1}$& - \\
   \hline
   $\delta$&$2\pi$/26& From Ref. [6]\\
   \hline
   $\delta_{0}$&$\delta$/2& -\\
   \hline
   $T_{0}$&295.85K& Room temperature\\
   \hline
   $\tau_{0}$&505pN-nm& Typical value\\
   \hline
   $\tau^{+}$&15$K_{B}$T/0.95$\delta$& Typical value\\
   \hline
   $\tau^{-}$&15$K_{B}$T/0.05$\delta$& Typical value\\
   \hline
   $k_{+}$(symm. case)&12000$s^{-1}$& Fitting data\\
   \hline
   $k_{-}$(symm. case)&2$k_{+}$& From theory\\
   \hline
   $k_{+}$(asymm. case)&10000$s^{-1}$& Fitting data\\
   \hline
   $k_{-}$(asymm. case)&20$k_{+}$& From theory\\
   \hline
   $\kappa_{0}$&400pN-nm-rad$^{-1}$& From Ref. [3]\\
   \hline
   $\kappa_{1}$&4000pN-nm-rad$^{-1}$& From Ref. [3]\\
   \hline
   $(\theta-\theta_{L})_{0}$&$2\pi/3$& From Ref. [3]\\
   \hline
   $\Delta(\theta-\theta_{L})$&2$\pi$/7& From Ref. [3]\\
   \hline
   $\Delta\theta_{0}$&$\delta_{0}$&  - \\
   \hline
\end{tabular}
\caption{Parameters used in the calculation.}
\end{table}
%\end{center}
\end{document}